\documentclass{article}

\usepackage[final, nonatbib]{neurips_mlsb_2021}

\usepackage[numbers]{natbib}

\usepackage[utf8]{inputenc} 
\usepackage[T1]{fontenc}    
\usepackage{hyperref}       
\usepackage{url}            
\usepackage{booktabs}       
\usepackage{amsfonts}       
\usepackage{nicefrac}       
\usepackage{microtype}      
\usepackage{xcolor}         
\usepackage{graphicx}
\usepackage{amsmath}

\title{MOLUCINATE: A Generative Model for Molecules in 3D Space}

\author{%
  Michael Arcidiacono \\
  \texttt{mixarcidiacono@gmail.com}
  \And
  David Ryan Koes \\
  Comp. \& Systems Biology \\
  University of Pittsburgh \\
  Pittsburgh, PA 15213 \\
  \texttt{dkoes@pitt.edu} \\
}
 
\begin{document}

\maketitle

\begin{abstract}

Recent advances in machine learning have enabled generative models for both optimization and \emph{de novo} generation of drug candidates with desired properties. Previous generative models have focused on producing SMILES strings or 2D molecular graphs, while attempts at producing molecules in 3D have focused on reinforcement learning (RL), distance matrices, and pure atom density grids. Here we present MOLUCINATE (MOLecUlar ConvolutIoNal generATive modEl), a novel architecture that simultaneously generates topological and 3D atom position information. We demonstrate the utility of this method by using it to optimize molecules for desired radius of gyration. In the future, this model can be used for more useful optimization such as binding affinity for a protein target.

\end{abstract}

\section{Introduction}

Computational drug discovery aims to generate molecules with optimized activity against protein targets. Given the success of machine learning (ML) techniques in image and text processing, there has been interest in applying similar techniques to \emph{de novo} drug design. Recent successes at generation of SMILES strings \cite{weininger_smiles_1988, gomez-bombarelli_automatic_2018, kusner_grammar_2017, hong_molecular_2019, maziarka_mol-cyclegan_2020} and graphs \cite{de_cao_molgan_2018, zang_moflow_2020, simonovsky_graphvae_2018, jin_junction_2019} are promising, but most do not generate 3D conformer information. Since a molecule's binding affinity against a protein target is dependant on its 3D conformation within the binding site, it is desirable to generate this information as well. \cite{wallach_atomnet_2015, ragoza_proteinligand_2017, jimenez_kdeep_2018, gomes_atomic_2017}. 

Current work in this area poses the problem of 3D molecular generation as a reinforcement learning (RL) problem \cite{gebauer_symmetry-adapted_2020, simm_3d_nodate, li_learning_2021, meldgaard_generating_2021}, which show promise for \emph{de novo} molecule generation. However, RL-based methods do not share certain advantages with generative methods that produce a latent space of molecules, allowing interpolation between molecules and optimization of properties directly via gradient descent. Among non-RL-based methods, there has been some work on generating distance matrices \cite{gebauer_generating_2018, hoffmann_generating_2019, nesterov_3dmolnet_nodate}, but this research is preliminary and so far restricted to very small molecules and constitutional isomers. Additionally, \citet{samanta_nevae_2019} generate 3D coordinates directly using graph convolutions, but this approach fails to encode inductive biases about 3D geometry and thus generates unrealistic coordinates. Lastly, \citet{ragoza_proteinligand_2017} generate atomic density grids, which has the advantage of being the same format as the input to convolutional networks predicting binding affinity \cite{wallach_atomnet_2015, ragoza_proteinligand_2017, jimenez_kdeep_2018, gomes_atomic_2017}, but requires an extra overhead for decoding the molecule topology from the grid. In this work, we build upon this representation to also incorporate information about the molecular graph. By simultaneously generating the molecular graph and its 3D structure, we can create a latent space of molecules that enables easy optimization of 3D properties while maintaining topological similarity to the original molecule. In the future, this model can be used for tasks such as \emph{de novo} drug discovery and optimization of candidate molecules for potency against a protein target.

In this work, we use a novel representation of 3D molecules that use per-atom voxel grids to encode atom position information. By combining spatial and graph convolutions into a novel architecture, we create a variational auto-encoder (VAE) \cite{kingma_auto-encoding_2014} to generate a latent space of 3D molecules. Additionally, we demonstrate the usefulness of this auto-encoder by using it to optimize molecules for radius of gyration (Rg).  This is an inherently 3D property of a molecule and can be readily and unambiguously computed for any molecule. In the future, we hope to use this architecture to optimize molecules for more interesting metrics (such as binding affinity against protein targets) and generate molecules conditional on a protein binding pocket.

\section{Methods}

\subsection{Representation of Molecules}

In order to represent molecules in 3D, we store each molecule as a graph $\mathcal{G}$ with a list of nodes $\mathcal{A}$ and edges $\mathcal{E}$. The atoms are ordered according to the canonical SMILES representation (as returned by RDKit \cite{noauthor_rdkit_nodate}). Each atom $a_i \in \mathcal{A}$ contains information about its type $t_i \in \{ \text{C}, \text{N}, \text{O}, \text{P}, \text{S}, \text{F}, \text{Cl}, \text{Br}, \text{I}, \text{END} \}$ and valence $v_i \in \mathbb{N}^3$. The atom valences tensor specifies the number of total bonds of each type (single, double triple) that the atom has to other (heavy) atoms. For instance, a carbon in a benzene ring would have its valence encoded as $[1, 1, 0]$. A final, virtual, atom of type $\text{END}$ is also appended to the graph.

To store 3D information, each atom has a position tensor with shape $(G, G, G)$, where $G$ is the grid size. For this application, an 8 \AA\ grid with 0.5 \AA\ resolution was used, so $G = 16$. We used two separate density encodings for the input and the output of the VAE. For the input molecules, the density function is taken from \citet{ragoza_proteinligand_2017} and is based on 3D exponential decay defined by the Van Der Waals radius. For the output, we one-hot encode the atom positions: we set the voxel that contains the atom center to 1 and set all other voxels to zero. While the one-hot representation is too sparse for the encoder's 3D convolutions \cite{kuzminykh_3d_2018}, it is useful for the output because it enables us to softmax the output ``densities'' over all space, allowing us to interpret the output as a 3D probability distribution.

For simplicity, we store bond types as per-atom information. Each atom $a_i$ has a bond type list $b_{ij} \in \{ \text{SINGLE}, \text{DOUBLE}, \text{TRIPLE}, \text{NO\_BOND} \}$ defined as the list of bond types to previous atoms $a_j$, where $j<i$. Each atom can have a maximum of 4 bonds to previous atoms and lists with less than 4 elements are padded with the extra bond type NO\_BOND. Note that, since double and triple bonds are encoded as individual ``bonds'' with different types, the 4 bond maximum still allows relevant hypervalent moieties such as phosphate groups.

\subsection{Combining Graph and Spatial Convolutions}

To encode and decode the atom position grids, graph and spatial convolutions were combined. For layer $l$ in the network, for each atom $i$, we define a voxel hidden state $h_i^{(l)}$ with shape $(C_l, G_l, G_l, G_l)$, where $C_l$ is the number of channels (the input atom positions have $C_0$ = 1) and $G_l$ is the grid size for that layer. To perform the graph-spatial convolution, the summed channels of each atom's neighbors are concatenated with the atom's channels, and then a 3x3x3 spatial convolution is performed.

\subsection{Bond prediction}

To predict bonded atoms, we take inspiration from the attention mechanism. For each atom $a_i$, we separately predict 4 bonds to atoms before in the range $[0, i)$. Once the VAE has decoded a hidden state $h_i$, for each bond type $b\in b_i$ we use a linear layer (followed by a leaky ReLU) to compute keys $k_i^{b}$. We compute values $v_i^{b}$ in the same way. The score for atom $a_i$ bonded to atom $a_{j<i}$ is simply $k_i^{b\top}v_j^{b}$. When decoding the final molecule, a bond is created between $a_i$ and the top scoring atom $a_j$ if $b \neq \text{NO\_BOND}$.

\subsection{VAE Architecture}

\begin{figure}
    \centering
    \includegraphics[scale=0.45]{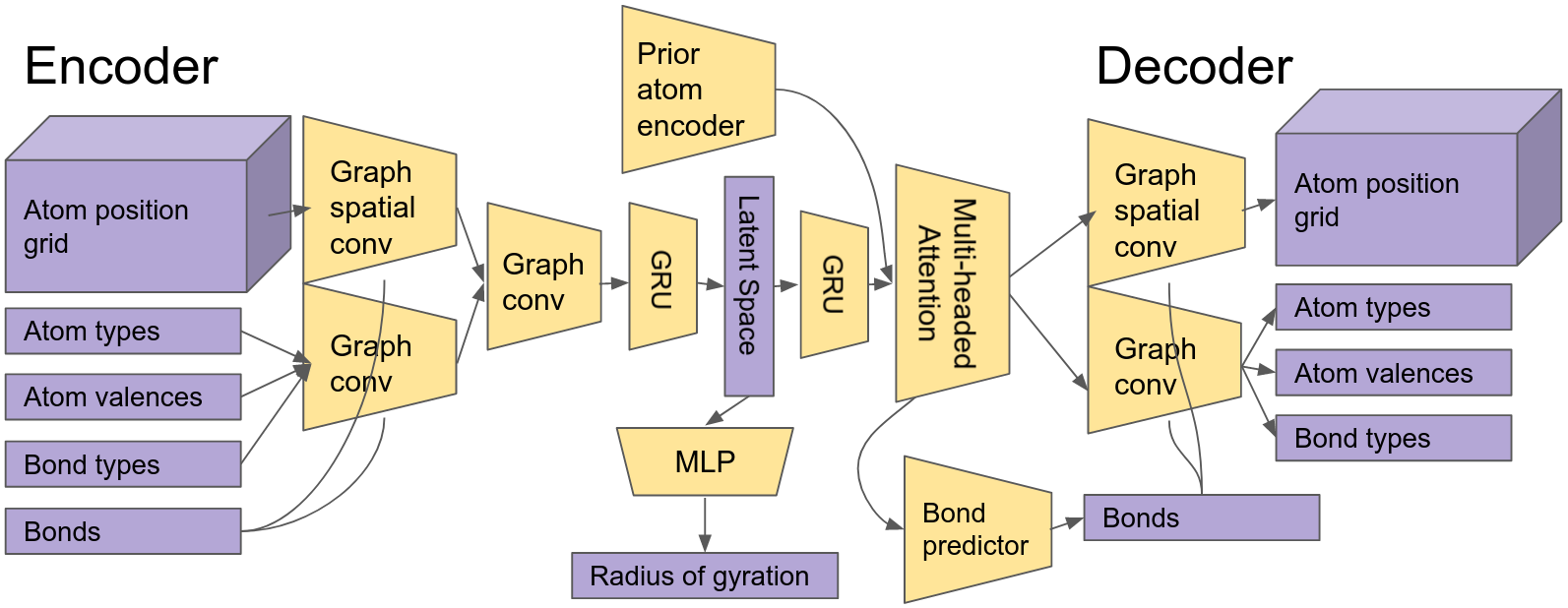}
    \caption{Model architecture.}
    \label{fig:model_arch}
\end{figure}

The overall auto-encoder architecture is shown in Figure \ref{fig:model_arch}. To encode a molecule into the latent space, we first encode the atom density grids into single per-atom vectors by alternating spatial-graph convolutional layers and 2x2x2 max pooling. This hidden state is then concatenated with the hidden state obtained by running graph convolutions on the remaining atom features (atom type, valences, and bonds types). After concatenation, further graph convolutions are used to produce the final per-atom hidden state. To turn this into a single latent code, a gated recurrent unit (GRU) \cite{cho_learning_2014} is used. The latent space size was 512.

To decode, we generate new atoms autoregressively. To generate atom $a_i$, we encode atoms $a_0, ..., a_{i-1}$ via an encoder with a similar architecture to the one described above -- however, all bonds to atoms $i$ or greater are masked out of the graph convolutions to prevent cheating. This encoded representation is concatenated with the latent code and a GRU is run to produce a new hidden representation. This is then fed into a multi-headed attention module \cite{vaswani_attention_2017} with 8 heads. To run graph convolutions, we need bonding information, so the bond predictor is immediately run from this hidden state. Using the predicted bonds, we run graph convolutions to produce the atom types, atom valences, and bond types. We additionally use spatial-graph convolutions to produce the predicted atom positions. To generate new molecules, we iteratively run the decoder to produce atoms one by one until an atom is reached with type $\text{END}$.

Additionally, we desire to predict Rg from the latent code. This is accomplished with a simple two-layer multi-layer perceptron (MLP). This property predictor is trained concurrently with the auto-encoder.

\subsection{Training}
\label{training}
The VAE was trained separately on the QM9 \cite{ruddigkeit_enumeration_2012, ramakrishnan_quantum_2014} and ZINC 15 datasets \cite{sterling_zinc_2015, irwin_zinc_2005}. The datasets were filtered down to molecules with 16 or fewer heavy atoms that fit into an 8 \AA\ grid (60,888 files for ZINC, 121,836 for QM9). The QM9 dataset was divided according to a 80:10:10 train:validation:test split, while ZINC was divided with a 90:10 train:validation split.\footnote{Due to an error early in training, a proper test set for ZINC was not used. No such mistake was made for QM9.} During training, molecules were loaded, kekulized, rotated randomly, and then fed into the VAE. The model was trained to simultaneously reconstruct the input and predict Rg. To do this, we utilized seven loss functions: the cross-entropy for the atom types, atom valences, atom positions, bond types, and bonded atoms, mean-squared-error (MSE) for Rg, and Kullback–Leibler (KL) divergence of the latent code to a multivariate normal distribution. The loss weights were, respectively, 1.0, 1.0, 0.6, 2.0, 2.0, 1.0, and 0.1. The VAE was optimized using AdamW \cite{loshchilov_decoupled_2019} with a learning rate of 0.001 and batch size of 64 for 168 epochs.

\section{Results}

The model was evaluated according to its ability to reconstruct, generate, and optimize molecules. To test reconstruction accuracy, molecules were sampled from the test dataset, encoded, and then decoded. The topological accuracy is the fraction of molecules whose decoded molecular graphs were exactly the same as the input molecules. Of the molecules that were correctly decoded, the RMSD of the conformers was also measured. On the QM9 test set, our model achieved a reconstruction accuracy of 94.7\% and the mean RMSD was 0.15 \AA. On the ZINC validation set, the Reconstruction accuracy was 50.6\% and the mean RMSD was 0.62 \AA.

To test generation, 16,000 latent points were randomly sampled from a multivariate normal distribution. The topological validity was defined to be the fraction of generated molecules that did not throw an exception when using RDKit's \verb|SanitizeMol|. Additionally, the stability of the generated conformers was tested by running Unified Force Field (UFF) optimization \cite{rappe_uff_1992}. The geometric validity was the fraction of topologically valid molecules that did not throw an exception when running UFF optimization. The RMSD between the (aligned) optimized conformer and the generated conformer was also measured for each geometrically valid molecule. Lastly, the distribution of computed properties of the generated molecules was compared to the distribution of the filtered original dataset. Of the molecules generated from the QM9 model, 84.0\% were topologically valid. Of those, 75.0\% were geometrically valid. Of the geometrically valid molecules, the mean UFF RMSD was 1.23 \AA. For the ZINC model, the topological validity was 60.3\%, the geometric validity was 75.0\%, and the UFF RMSD was 1.05 \AA.

\begin{figure}
    \centering
    \includegraphics[scale=0.15]{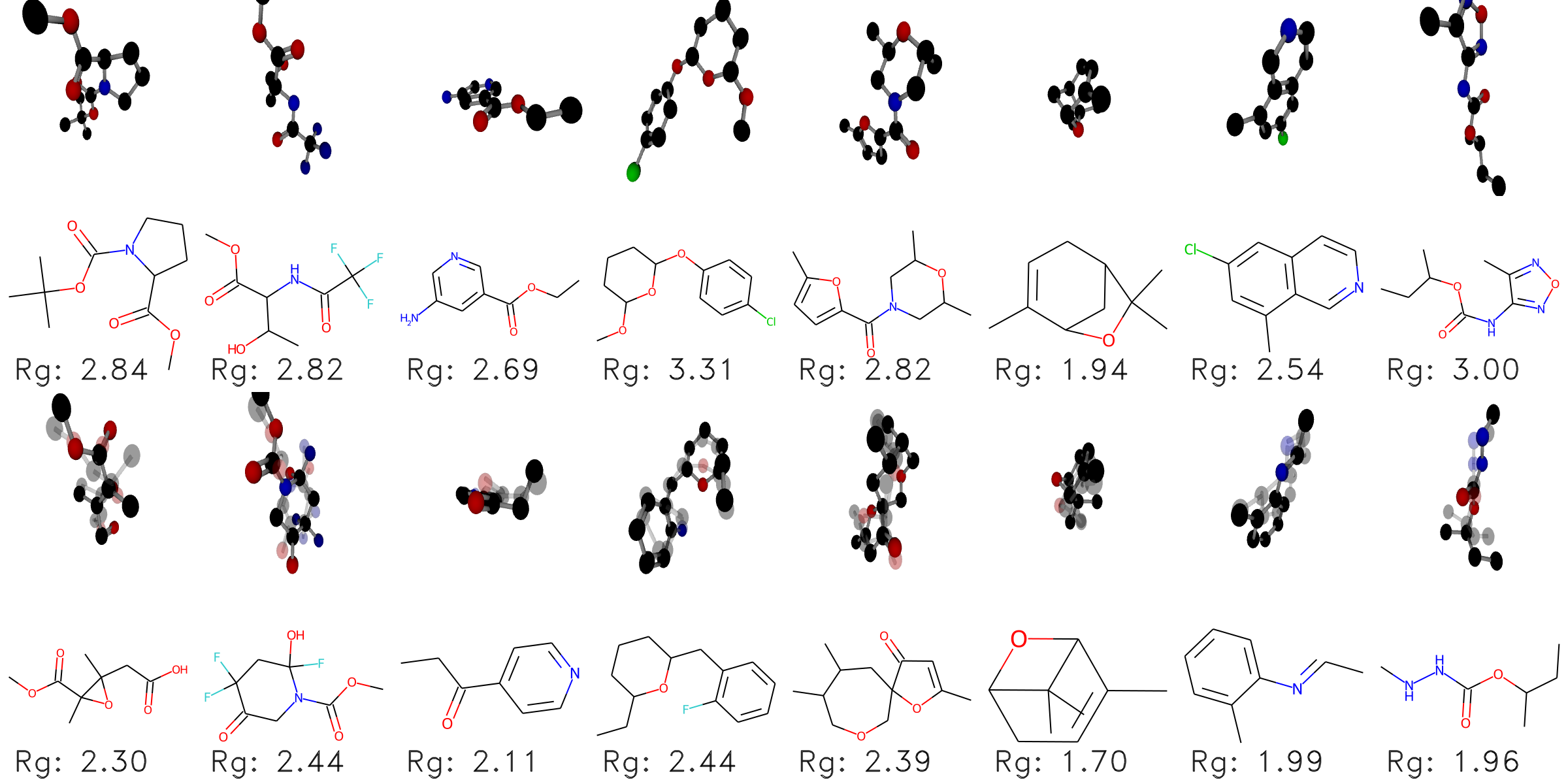}
    \caption{Results of minimizing Rg of molecules sampled from the ZINC validation set. The top row shows the original molecules, and the bottom row shows the molecules after optimization. The optimized molecules are shown after UFF relaxation; the raw output of the model is overlaid transparently.}
    \label{fig:optim_small}
\end{figure}

To test optimization, molecules were chosen from the ZINC validation set and encoded into latent vectors. Rg was predicted using the prediction network on the latent vector, and the derivative of the predicted radius with respect to the latent vector was computed. Gradient descent was then run on the latent vector for 500 iterations (learning rate 0.2) to minimize the radius. The results of Rg minimization are shown in Figure \ref{fig:optim_small}. After optimizing 1,000 molecules, 75.0\% of the resulting molecules were valid, 99.5\% of the valid molecules had reduced Rg, and the average difference in Rg was -0.53 \AA. Qualitatively, the results are consistent with the algorithm ``shrinking'' the molecules while preserving overall shape and features.


\section{Conclusions}
\label{conclusion}

By using a novel graph representation of molecules that incorporates voxelized atom position information, we have created a generative model that allows encoding, decoding, interpolation, and optimization of both the geometric and topological properties of molecules. Our model produces topologically valid molecules 60.3\% of the time on the ZINC validation set, and the UFF RMSD of 1.0 \AA\ demonstrates that the molecules produced are often close to low energy conformations. However, the fact that the UFF optimization routine failed to converge for a quarter of valid molecules implies that the actual mean RMSD to a low-energy structure might be larger. More tweaking of the UFF optimization parameters needs to be done so that it does not fail so often.

We have also demonstrated that this model can optimize molecules for characteristics defined by the molecule's 3D geometry -- here, Rg. While this is admittedly a toy problem, the optimization serves as a useful proof of concept. In the future, we hope to use this model to optimize molecules for more useful characteristics such as binding affinity for protein targets.

Perhaps the biggest current limitation of the model was the small size of molecules used: the model was only trained and tested on molecules with 16 or fewer heavy atoms that fit into an 8 \AA\ cube. These parameters were chosen for ease of iteration; since the model run time scales as $O(ag^3)$, where $a$ is the number of atoms and $g$ is the grid size, supporting larger molecules takes more time to train. Now that we have narrowed in on an architecture that works for these small molecules, we hope to scale up the model to support larger and more useful molecules.

The full source code for training and evaluating this model is available at \url{https://github.com/mixarcid/molucinate}. 

\section{Acknowledgements}
The authors thank Dr. Alexander Tropsha, Matthew Ragoza, and Jack Lynch for their insightful comments while discussing this work. Additionally, this work is supported by R01GM108340 from the National Institute of General Medical Sciences.

\bibliographystyle{plainnat}
\bibliography{main}

\begin{thebibliography}{32}
\providecommand{\natexlab}[1]{#1}
\providecommand{\url}[1]{\texttt{#1}}
\expandafter\ifx\csname urlstyle\endcsname\relax
  \providecommand{\doi}[1]{doi: #1}\else
  \providecommand{\doi}{doi: \begingroup \urlstyle{rm}\Url}\fi

\bibitem[noa()]{noauthor_rdkit_nodate}
{RDKit}: Open-source cheminformatics.
\newblock URL \url{https://www.rdkit.org/}.

\bibitem[Cho et~al.()Cho, van Merrienboer, Gulcehre, Bahdanau, Bougares,
  Schwenk, and Bengio]{cho_learning_2014}
Kyunghyun Cho, Bart van Merrienboer, Caglar Gulcehre, Dzmitry Bahdanau, Fethi
  Bougares, Holger Schwenk, and Yoshua Bengio.
\newblock Learning phrase representations using {RNN} encoder-decoder for
  statistical machine translation.
\newblock URL \url{http://arxiv.org/abs/1406.1078}.

\bibitem[De~Cao and Kipf()]{de_cao_molgan_2018}
Nicola De~Cao and Thomas Kipf.
\newblock {MolGAN}: An implicit generative model for small molecular graphs.
\newblock URL \url{http://arxiv.org/abs/1805.11973}.

\bibitem[Gebauer et~al.({\natexlab{a}})Gebauer, Gastegger, and
  Schütt]{gebauer_generating_2018}
Niklas W.~A. Gebauer, Michael Gastegger, and Kristof~T. Schütt.
\newblock Generating equilibrium molecules with deep neural networks.
\newblock {\natexlab{a}}.
\newblock URL \url{http://arxiv.org/abs/1810.11347}.

\bibitem[Gebauer et~al.({\natexlab{b}})Gebauer, Gastegger, and
  Schütt]{gebauer_symmetry-adapted_2020}
Niklas W.~A. Gebauer, Michael Gastegger, and Kristof~T. Schütt.
\newblock Symmetry-adapted generation of 3d point sets for the targeted
  discovery of molecules.
\newblock {\natexlab{b}}.
\newblock URL \url{http://arxiv.org/abs/1906.00957}.

\bibitem[Gomes et~al.()Gomes, Ramsundar, Feinberg, and
  Pande]{gomes_atomic_2017}
Joseph Gomes, Bharath Ramsundar, Evan~N. Feinberg, and Vijay~S. Pande.
\newblock Atomic convolutional networks for predicting protein-ligand binding
  affinity.
\newblock URL \url{http://arxiv.org/abs/1703.10603}.

\bibitem[Gómez-Bombarelli et~al.()Gómez-Bombarelli, Wei, Duvenaud,
  Hernández-Lobato, Sánchez-Lengeling, Sheberla, Aguilera-Iparraguirre,
  Hirzel, Adams, and Aspuru-Guzik]{gomez-bombarelli_automatic_2018}
Rafael Gómez-Bombarelli, Jennifer~N. Wei, David Duvenaud, José~Miguel
  Hernández-Lobato, Benjamín Sánchez-Lengeling, Dennis Sheberla, Jorge
  Aguilera-Iparraguirre, Timothy~D. Hirzel, Ryan~P. Adams, and Alán
  Aspuru-Guzik.
\newblock Automatic chemical design using a data-driven continuous
  representation of molecules.
\newblock 4\penalty0 (2):\penalty0 268--276.
\newblock ISSN 2374-7943, 2374-7951.
\newblock \doi{10.1021/acscentsci.7b00572}.
\newblock URL \url{http://arxiv.org/abs/1610.02415}.

\bibitem[Hoffmann and Noé()]{hoffmann_generating_2019}
Moritz Hoffmann and Frank Noé.
\newblock Generating valid euclidean distance matrices.
\newblock URL \url{http://arxiv.org/abs/1910.03131}.

\bibitem[Hong et~al.()Hong, Lim, Ryu, and Kim]{hong_molecular_2019}
Seung~Hwan Hong, Jaechang Lim, Seongok Ryu, and Woo~Youn Kim.
\newblock Molecular generative model based on adversarially regularized
  autoencoder.
\newblock URL \url{http://arxiv.org/abs/1912.05617}.

\bibitem[Irwin and Shoichet()]{irwin_zinc_2005}
John~J. Irwin and Brian~K. Shoichet.
\newblock {ZINC} – a free database of commercially available compounds for
  virtual screening.
\newblock 45\penalty0 (1):\penalty0 177--182.
\newblock ISSN 1549-9596.
\newblock \doi{10.1021/ci049714}.
\newblock URL \url{https://www.ncbi.nlm.nih.gov/pmc/articles/PMC1360656/}.

\bibitem[Jiménez et~al.()Jiménez, Škalič, Martínez-Rosell, and
  De~Fabritiis]{jimenez_kdeep_2018}
José Jiménez, Miha Škalič, Gerard Martínez-Rosell, and Gianni
  De~Fabritiis.
\newblock {KDEEP}: Protein–ligand absolute binding affinity prediction via
  3d-convolutional neural networks.
\newblock 58\penalty0 (2):\penalty0 287--296.
\newblock ISSN 1549-9596.
\newblock \doi{10.1021/acs.jcim.7b00650}.
\newblock URL \url{https://doi.org/10.1021/acs.jcim.7b00650}.
\newblock Publisher: American Chemical Society.

\bibitem[Jin et~al.()Jin, Barzilay, and Jaakkola]{jin_junction_2019}
Wengong Jin, Regina Barzilay, and Tommi Jaakkola.
\newblock Junction tree variational autoencoder for molecular graph generation.
\newblock URL \url{http://arxiv.org/abs/1802.04364}.

\bibitem[Kingma and Welling()]{kingma_auto-encoding_2014}
Diederik~P. Kingma and Max Welling.
\newblock Auto-encoding variational bayes.
\newblock URL \url{http://arxiv.org/abs/1312.6114}.

\bibitem[Kusner et~al.()Kusner, Paige, and
  Hernández-Lobato]{kusner_grammar_2017}
Matt~J. Kusner, Brooks Paige, and José~Miguel Hernández-Lobato.
\newblock Grammar variational autoencoder.
\newblock URL \url{http://arxiv.org/abs/1703.01925}.

\bibitem[Kuzminykh et~al.()Kuzminykh, Polykovskiy, Kadurin, Zhebrak, Baskov,
  Nikolenko, Shayakhmetov, and Zhavoronkov]{kuzminykh_3d_2018}
Denis Kuzminykh, Daniil Polykovskiy, Artur Kadurin, Alexander Zhebrak, Ivan
  Baskov, Sergey Nikolenko, Rim Shayakhmetov, and Alex Zhavoronkov.
\newblock 3d molecular representations based on the wave transform for
  convolutional neural networks.
\newblock 15\penalty0 (10):\penalty0 4378--4385.
\newblock ISSN 1543-8384.
\newblock \doi{10.1021/acs.molpharmaceut.7b01134}.
\newblock URL \url{https://doi.org/10.1021/acs.molpharmaceut.7b01134}.
\newblock Publisher: American Chemical Society.

\bibitem[Li et~al.()Li, Pei, and Lai]{li_learning_2021}
Yibo Li, Jianfeng Pei, and Luhua Lai.
\newblock Learning to design drug-like molecules in three-dimensional space
  using deep generative models.
\newblock URL \url{http://arxiv.org/abs/2104.08474}.

\bibitem[Loshchilov and Hutter()]{loshchilov_decoupled_2019}
Ilya Loshchilov and Frank Hutter.
\newblock Decoupled weight decay regularization.
\newblock URL \url{http://arxiv.org/abs/1711.05101}.

\bibitem[Maziarka et~al.()Maziarka, Pocha, Kaczmarczyk, Rataj, Danel, and
  Warchoł]{maziarka_mol-cyclegan_2020}
Łukasz Maziarka, Agnieszka Pocha, Jan Kaczmarczyk, Krzysztof Rataj, Tomasz
  Danel, and Michał Warchoł.
\newblock Mol-{CycleGAN}: a generative model for molecular optimization.
\newblock 12\penalty0 (1):\penalty0 2.
\newblock ISSN 1758-2946.
\newblock \doi{10.1186/s13321-019-0404-1}.
\newblock URL \url{https://doi.org/10.1186/s13321-019-0404-1}.

\bibitem[Meldgaard et~al.()Meldgaard, Köhler, Mortensen, Christiansen, Noé,
  and Hammer]{meldgaard_generating_2021}
Søren~Ager Meldgaard, Jonas Köhler, Henrik~Lund Mortensen, Mads-Peter~V.
  Christiansen, Frank Noé, and Bjørk Hammer.
\newblock Generating stable molecules using imitation and reinforcement
  learning.
\newblock URL \url{http://arxiv.org/abs/2107.05007}.

\bibitem[Nesterov et~al.()Nesterov, Wieser, and Roth]{nesterov_3dmolnet_nodate}
Vitali Nesterov, Mario Wieser, and Volker Roth.
\newblock 3dmolnet: A generative network for molecular structures.
\newblock page~5.

\bibitem[Ragoza et~al.()Ragoza, Hochuli, Idrobo, Sunseri, and
  Koes]{ragoza_proteinligand_2017}
Matthew Ragoza, Joshua Hochuli, Elisa Idrobo, Jocelyn Sunseri, and David~Ryan
  Koes.
\newblock Protein–ligand scoring with convolutional neural networks.
\newblock 57\penalty0 (4):\penalty0 942--957.
\newblock ISSN 1549-9596.
\newblock \doi{10.1021/acs.jcim.6b00740}.
\newblock URL \url{https://doi.org/10.1021/acs.jcim.6b00740}.
\newblock Publisher: American Chemical Society.

\bibitem[Ramakrishnan et~al.()Ramakrishnan, Dral, Rupp, and von
  Lilienfeld]{ramakrishnan_quantum_2014}
Raghunathan Ramakrishnan, Pavlo~O. Dral, Matthias Rupp, and O.~Anatole von
  Lilienfeld.
\newblock Quantum chemistry structures and properties of 134 kilo molecules.
\newblock 1\penalty0 (1):\penalty0 140022.
\newblock ISSN 2052-4463.
\newblock \doi{10.1038/sdata.2014.22}.
\newblock URL \url{https://www.nature.com/articles/sdata201422}.
\newblock Bandiera\_abtest: a Cg\_type: Nature Research Journals Number: 1
  Primary\_atype: Research Publisher: Nature Publishing Group Subject\_term:
  Computational chemistry;Density functional theory;Quantum chemistry
  Subject\_term\_id:
  computational-chemistry;density-functional-theory;quantum-chemistry.

\bibitem[Rappe et~al.()Rappe, Casewit, Colwell, Goddard, and
  Skiff]{rappe_uff_1992}
A.~K. Rappe, C.~J. Casewit, K.~S. Colwell, W.~A. Goddard, and W.~M. Skiff.
\newblock {UFF}, a full periodic table force field for molecular mechanics and
  molecular dynamics simulations.
\newblock 114\penalty0 (25):\penalty0 10024--10035.
\newblock ISSN 0002-7863.
\newblock \doi{10.1021/ja00051a040}.
\newblock URL \url{https://doi.org/10.1021/ja00051a040}.
\newblock Publisher: American Chemical Society.

\bibitem[Ruddigkeit et~al.()Ruddigkeit, van Deursen, Blum, and
  Reymond]{ruddigkeit_enumeration_2012}
Lars Ruddigkeit, Ruud van Deursen, Lorenz~C. Blum, and Jean-Louis Reymond.
\newblock Enumeration of 166 billion organic small molecules in the chemical
  universe database {GDB}-17.
\newblock 52\penalty0 (11):\penalty0 2864--2875.
\newblock ISSN 1549-9596.
\newblock \doi{10.1021/ci300415d}.
\newblock URL \url{https://doi.org/10.1021/ci300415d}.
\newblock Publisher: American Chemical Society.

\bibitem[Samanta et~al.()Samanta, De, Jana, Chattaraj, Ganguly, and
  Gomez-Rodriguez]{samanta_nevae_2019}
Bidisha Samanta, Abir De, Gourhari Jana, Pratim~Kumar Chattaraj, Niloy Ganguly,
  and Manuel Gomez-Rodriguez.
\newblock {NeVAE}: A deep generative model for molecular graphs.
\newblock URL \url{http://arxiv.org/abs/1802.05283}.

\bibitem[Simm et~al.()Simm, Pinsler, Csányi, and
  Hernández-Lobato]{simm_3d_nodate}
Gregor N~C Simm, Robert Pinsler, Gábor Csányi, and José~Miguel
  Hernández-Lobato.
\newblock 3d molecular design with covariant neural networks.
\newblock page~6.

\bibitem[Simonovsky and Komodakis()]{simonovsky_graphvae_2018}
Martin Simonovsky and Nikos Komodakis.
\newblock {GraphVAE}: Towards generation of small graphs using variational
  autoencoders.
\newblock URL \url{http://arxiv.org/abs/1802.03480}.

\bibitem[Sterling and Irwin()]{sterling_zinc_2015}
Teague Sterling and John~J. Irwin.
\newblock {ZINC} 15 – ligand discovery for everyone.
\newblock 55\penalty0 (11):\penalty0 2324--2337.
\newblock ISSN 1549-9596.
\newblock \doi{10.1021/acs.jcim.5b00559}.
\newblock URL \url{https://doi.org/10.1021/acs.jcim.5b00559}.
\newblock Publisher: American Chemical Society.

\bibitem[Vaswani et~al.()Vaswani, Shazeer, Parmar, Uszkoreit, Jones, Gomez,
  Kaiser, and Polosukhin]{vaswani_attention_2017}
Ashish Vaswani, Noam Shazeer, Niki Parmar, Jakob Uszkoreit, Llion Jones,
  Aidan~N. Gomez, Lukasz Kaiser, and Illia Polosukhin.
\newblock Attention is all you need.
\newblock URL \url{http://arxiv.org/abs/1706.03762}.

\bibitem[Wallach et~al.()Wallach, Dzamba, and Heifets]{wallach_atomnet_2015}
Izhar Wallach, Michael Dzamba, and Abraham Heifets.
\newblock {AtomNet}: A deep convolutional neural network for bioactivity
  prediction in structure-based drug discovery.
\newblock URL \url{http://arxiv.org/abs/1510.02855}.

\bibitem[Weininger()]{weininger_smiles_1988}
David Weininger.
\newblock {SMILES}, a chemical language and information system. 1. introduction
  to methodology and encoding rules.
\newblock 28\penalty0 (1):\penalty0 31--36.
\newblock ISSN 0095-2338.
\newblock \doi{10.1021/ci00057a005}.
\newblock URL \url{https://pubs.acs.org/doi/abs/10.1021/ci00057a005}.
\newblock Publisher: American Chemical Society.

\bibitem[Zang and Wang()]{zang_moflow_2020}
Chengxi Zang and Fei Wang.
\newblock {MoFlow}: An invertible flow model for generating molecular graphs.
\newblock pages 617--626.
\newblock \doi{10.1145/3394486.3403104}.
\newblock URL \url{http://arxiv.org/abs/2006.10137}.

\end{thebibliography}

\appendix




\end{document}